\documentclass[english,twocolumn,preprintnumbers,amsmath,amssymb]{revtex4}
\usepackage[T1]{fontenc}
\usepackage[latin9]{inputenc}
\usepackage{amsmath}
\usepackage{graphicx}

\makeatletter

\providecommand{\tabularnewline}{\\}

\makeatletter

\usepackage{babel}
\makeatother

\begin{document}

\title{Emergence of symmetry from random $n$-body interactions}

\author{Alexander Volya}

\affiliation{Department of Physics, Florida State University, Tallahassee, FL
32306-4350, USA}

\date{\today}

\begin{abstract}
An ensemble with random $n$-body interactions is investigated in
the presence of symmetries. A striking emergence of regularities in
spectra, ground state spins and isospins is discovered in both odd
and even-particle systems. Various types of correlations from pairing
to spectral sequences and correlations across different masses are
explored. A search for interpretation is presented. 
\end{abstract}



\maketitle
Recent progress in \emph{ab-initio} treatment of light nuclei show
unambiguously the essential role played by the three and four-body
forces \citep{Navratil:2002,Pieper:2001}. With the advancement of
the mesoscopic physics and with the ability to manipulate interactions
the question how $n$-body interactions play out in the many-body
physics becomes increasingly important. 

It is known since the Wigner-Dyson random matrix theory (RMT) \citep{Porter:1965},
see also \citep{Guhr:1998,Brody:1981}, that complex configuration
mixing driven by many-body forces have generic statistical features
which also depend on the nature of interaction \citep{French:1971}.
Symmetries have a robust effect, which is not fully understood. Recently,
a remarkable geometrical ordering have been numerically found in the
Two-Body Random Ensembles (2-BRE)\citep{Johnson:1998}. The main question
is centered around the disproportionally large probability for the
ground state (g.s.) spin $J$ to be zero, $J_{0}=0$, in the even-particle
system. Manifestations of the symmetry were seen in features of the
mean-field \citep{ZELEVINSKY:1993,Bijker:2002}, and in transition
probabilities indicating vibrational and rotational low-lying spectra
\citep{Zelevinsky:2004}. Starting from the pioneering paper \citep{Johnson:1998}
over a hundred works were published by different groups striving to
understand the emergence of symmetries, their role and origin in the
2-BRE. The summary of these efforts may be found in reviews \citep{Papenbrock:2007,Zhao:2004,Zelevinsky:2004}.
The success is mixed, we understand that pairing \citep{Mulhall:2000}
and time-reversal \citep{Bijker:1999} are not the primary causes.
The boson approximation of fermion pairs and chaos arising from complex
geometrical couplings, the \emph{geometric chaoticity}, provide some
qualitative understanding of the trend; more quantitative explanations
have been suggested in Refs. \citep{Zhao:2002,Papenbrock:2004} via
numerical observations of geometrical correlations and diagonalization
of the individual interaction terms. Nevertheless, the simple question
of symmetry and chaos asks for a simple answer which is still missing.
Driven by the quest for understanding of interplay between symmetry
and many-body complexity and the general interest and importance of
many-body interactions in modern physics in this work we address the
$n$-body Random Ensembles ($n-$BRE) with symmetries. The evolution
of spectrum and level spacing for the Gaussian $n$-BRE without symmetries
is discussed in Ref. \citep{French:1971}.

The $n$-body rotationally invariant interaction Hamiltonian is defined
as \[
H^{(n)}=\sum_{\alpha\beta}\sum_{L}V_{L}^{(n)}(\alpha\beta)\sum_{M=-L}^{L}\, T_{LM}^{(n)^{\dagger}}(\alpha)\, T_{LM}^{(n)}(\beta),\]
 where operators $T_{LM}^{(n)^{\dagger}}(\alpha)$ are $n$-body creation
operators coupled to a total angular momentum $L$ and magnetic projection
$M$, $T_{LM}^{(n)^{\dagger}}(\alpha)=\sum_{12\dots n}C_{12\dots n}^{LM}(\alpha)\, a_{1}^{\dagger}a_{2}^{\dagger}\dots a_{n}^{\dagger},$
here $1$ is the single-particle index. For $n=2$ the coefficients
$C_{12}^{LM}$ are proportional to the Clebsch-Gordan coefficients
and index $\alpha$ is uniquely defined by single-particle levels
involved. For $n>2$ the index $\alpha$ must include additional information
about the coupling scheme. Here we define the basis set of normalized
$n$-body operators $T_{LM}^{(n)}(\alpha)$ from a full set of orthogonal
$n$-body eigenstates $T_{LM}^{(n)^{\dagger}}(\alpha)|0\rangle$ of
an arbitrarily chosen reference two-body Hamiltonian $H_{0}^{(2)},$
solved numerically. The exact form of this Hamiltonian is irrelevant
as long as it preserves rotational and other symmetries of the problem. 

The $n$-BRE is defined as a set of $n$-body interaction Hamiltonians
that for an $n$-particle system leads to a Gaussian Orthogonal Ensemble
(GOE) within every symmetry class. The interaction strengths $V_{L}^{(n)}(\alpha,\beta)$
are selected at random with normal distribution centered at zero $\langle V_{L}^{(n)}(\alpha,\beta)\rangle=0$
and unit diagonal variance $\langle V_{L}^{(n)}(\alpha,\beta)\, V_{L'}^{(n)}(\alpha',\beta')\rangle=\delta_{LL'}\delta_{\alpha\alpha'}\delta_{\beta\beta'}(1+\delta_{\alpha\beta})/2.$
The time reversal symmetry sets $V_{L}^{(n)}(\alpha,\beta)=V_{L}^{(n)}(\beta,\alpha)$,
thus we assume ordered labels $\alpha\ge\beta.$ The ensemble is $H_{0}^{(2)}$
independent. The variance defines the energy unit. Our ensemble extends
the $n$-body Embedded GOE \citep{Mon:1975,Kota:2001} by incorporating
symmetries. Inclusion of the isospin symmetry is straightforward. 

Through the text we use $P\left[\text{event}\right]$ to denote chances
of observing a certain event in the $n$-BRE. Most commonly we discuss
$J(N)_{0}$ which is an event where g.s. of an $N$-particle system
has spin $J$; the subscript reflects the order in excitation energy,
0-for g.s. 1-first excited state, and etc. Where obvious we omit the
explicit reference to the particle number $N.$

\begin{figure}
\includegraphics[width=3.4in]{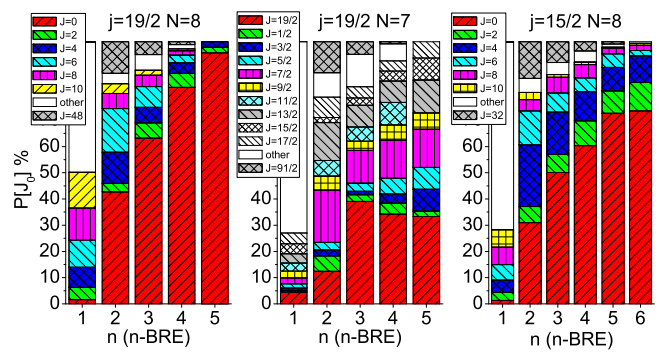}

\caption{(Color online) The stacked-bar histogram showing the distribution
of ground states with a given angular momentum $J$ for a single $j$-system,
with $j$ and $N$ as marked. The histograms are shown as a function
of $n$ for different $n$-BRE. For the $n=1$ case we assign probability
proportionally to the number of states with that spin. The stacking
order is marked, and corresponds to the increasing $J$ starting from
$J_{min}$, with the exception of the odd-$N$ where $J_{min}\equiv j.$
The spins $J$ with $P\left[J_{0}\right]<2\%$ are not separately
identified, their cumulative probability is shown with the white bar,
labeled {}``other''. \label{fig:j19N87}}

\end{figure}

We start with a single $j$-level model of $2j+1$ degeneracy with
$N$ identical fermions. In Fig.\,\ref{fig:j19N87} some examples
are shown. In the special case of $n=1$, the mean-field only, all
many-body states are degenerate. Here by definition we assume an equiprobable
ordering of the states in the spectrum. Thus, the probability to observe
a g.s. with a spin $J$, $P\left[J_{0}\right]$ is by definition proportional
to the number of many-body states with this spin in the model space
$d(J).$ Following the semi-classical consideration of the random-walk-type
vector couplings, see also \citep{Bethe:1936} $d(J)\sim(2J+1)\exp[-3J(J+1)/2Nj(j+1)]$
which disfavors both $J=0$ and the maximum possible spin $J_{max}=N(2j+1-N)/2$.
The equiprobable ordering is in drastic contrast to the 2-BRE where
g.s. is most likely to be $J_{0}=0$ and the probability for $J_{0}=J_{max}$
is large, Fig.\,\ref{fig:j19N87}. The preponderance of $P\left[0_{0}\right]$
becomes stronger for the $n$-BRE with higher $n$, and at the same
time the chances of $J_{max}$ as g.s. quickly diminish. For $n=4$
and 5, apart from the dominant $J_{0}=0$, the states with spins $J_{0}=2,4,6,8$
have small, few percent-level, chances to appear as g.s.

The middle panel in Fig.\,\ref{fig:j19N87} corresponds to the $N=7$
odd-particle system. The lowest bar in the stacked histogram shows
the $P\left[j_{0}\right]$ (where $j=19/2$) that can be interpreted
as a single particle coupled to the $J=0$ $N=6$ core. While for
the 2-BRE $P\left[19/2_{0}\right]=12.4(4)$\% (note that $P\left[13/2_{0}\right]=$14.5(4)\%),
the $P\left[19/2_{0}\right]$ is bigger for $n=3,\,4$ and 5. The
maximum spin probability $P\left[(91/2)_{0}\right]$ is enhanced for
the 2-BRE but similarly to the even-particle system declines for larger
$n.$ Here, and on some occasions below we include statistical errors
due to a limited number of random realizations tested.

An important case of $n=N$ is not show in Fig.\,\ref{fig:j19N87}
because of its drastic contrast to the $n<N$ situations. Here, within
each symmetry class the Hamiltonian matrix is represented by the GOE
for which with increasing dimensionality $d(J)$ the distribution
of eigenvalues quickly converges to the Wigner semicircle with the
radius $\sqrt{2d(J)}$. A detailed quantitative analysis can be done
using the RMT but it is clear that the probability $P\left[J_{0}\right]$
strongly favors those spins $J$ with the highest dimensionalities
$d(J)$. For example, for $N=8$ $j=19/2$ the highest dimensions
are 179, 178, 173, and 169 for spins $J=$12, 14, 16, and 10; the
corresponding probabilities $P\left[J_{0}\right]$ in the 8-BRE are
41, 35, 13 and 5\%, respectively. For lower $n=N-1$ the change in
the spin statistics of g.s. occurs abruptly. 

\begin{figure}
\includegraphics[width=3.4in]{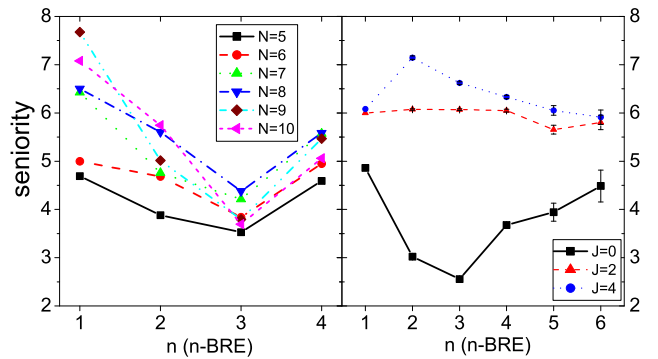}

\caption{The ensemble averaged seniority $s$ is shown as a function of $n$
for the $n$-BRE. Left panel corresponds to $j=19/2$ with connected
by line points for different particle numbers. Only $\left(J_{min}\right)_{0}$
ground state realizations are selected, namely $J_{0}=0$ for $N$
even and $J_{0}=19/2$ for odd. On the right the half-occupied system
$N=8$ $j=15/2$ is examined, which shows the average seniority in
realizations with g.s $J_{0}=0,2$ and 4. For the $n=1$ case, (equiprobable
ground state distribution) the average seniority of states with a
given spin is quoted. \label{fig:pairing}}

\end{figure}

Although pairing was ruled out as an explanation for the 2-BRE given
its dominance in realistic systems it is worth revisiting this question
for the $n$-BRE. In Fig.\,\ref{fig:pairing} the evolution of seniority
$s$ (the number of unpaired particles) as a function of $n$ is shown.
The seniority $s$ is evaluated using expectation value in the state
of interest $|N,\alpha\rangle$ of the $L=0$ pair operator, which
is unique in a single-$j$ $\langle N,\alpha|T_{00}^{(2)^{\dagger}}T_{00}^{(2)}|N,\alpha\rangle=(N-s)(2j+3-N-s)/\left\{ 2(2j+1)\right\} .$

Surprisingly, the paired state is favored by the three-body forces.
Significant lowering of the seniority for $n=3$ is a robust result
in all cases considered Fig.\,\ref{fig:pairing}. Given that the
$P[0_{0}]$ goes up for higher $n$ while $s$ is increasing for $n>3$
pairing still does not explain the preponderance of the zero spin
g.s. The structure of the two-body forces on a single-$j$ level is
known to facilitate seniority conservation, only about a third of
the $j-1/2$ independent linear combinations of interaction parameters
$V_{L}^{(2)}$ mix seniorities, seniority is a good quantum number
for any interaction on $j<9/2$ \citep{Volya:2002PRC}. There is lowering
of seniority against the average ($n=1$) for $n=2$ and perhaps explanation
for $n=3$ pairing enhancement lies in a similar seniority conserving
structure of the three-body interactions. This question is a subject
for future work.

The particle number $N$ and the Casimir operators for the symmetry
groups such as $J^{2}$ and $T^{2}$ are conserved quantities, and
in each random realization of the Hamiltonian are expected to describe
its coherent part. In the 2-BRE the particle-hole transformation can
be used to obtain a coherent $J^{2}$-dependent component of the Hamiltonian
$\langle H^{(2)}\rangle_{J}=\tilde{V}_{1}J^{2}.$ The moment of inertia
coefficient $\tilde{V_{1}}$ is given via angular momentum recoupling
coefficients, for a single-$j$ it is

\begin{equation}
\tilde{V}_{1}=\sum_{L}\frac{3(2L+1)}{j(j+1)(2j+1)}\left\{ \begin{array}{ccc}
j & j & 1\\
j & j & L\end{array}\right\} V_{L}^{(2)}.\label{eq:minertia}\end{equation}
 The 2-BRE g.s. spin systematics follows from here; see also geometric
chaoticity arguments \citep{Mulhall:2000}. The $\tilde{V}_{1}$,
being a sum of random numbers with normal distribution centered at
zero, is itself a normally distributed random variable with equal
chances of being positive and negative. Thus the $P[0_{0}]=P[(J_{max})_{0}]=1/2.$
This prediction qualitatively describes the observations, although
it is distorted by other incoherent interaction terms .

The emergence of the moment of inertia in each realization of interaction
is further supported by the mass number independence of the 2-BRE
results. Indeed, the moment of inertia \eqref{eq:minertia} is particle-number
independent, thus for a given realization of the two-body Hamiltonian
if $\tilde{V_{1}}>0$ the g.s. spin is zero $J(N)_{0}=0$ for any
even $N$ ($j$ for odd); similarly for $\tilde{V}_{1}<0$ the g.s.
is $J_{max}(N).$ To quantify the correlation between the statistics
for different $N$ we consider the joint probability that all systems
from $N=5$ to half-occupied, simultaneously have the maximum spin
$P\left[J_{max}(5)_{0},J_{max}(6)_{0},...\right]$. The same can be
done for the minimum spin $J_{min}$ which we define to be zero for
an even $N$ and s.p. $j$ for odd. Because of the particle-hole symmetry
the eigenvalues of the $2j+1-N$ system apart from a monopole constant
shift in energy are identical to the $N$-particle system. In an uncorrelated
case the joint probability is a product of the independent probabilities
$P\left[J(N_{1})_{0},J(N_{2})_{0},\dots\right]=\prod_{i}P\left[J(N_{i})_{0}\right]$
which is generally small. On the other hand, if strong correlations
exist the joint probability is of the order of individual $P\left[J_{0}\right]$.
In our example $j=19/2$ with $N=5,6\dots10$ $P\left[J_{max}(5)_{0}\dots J_{max}(10)_{0}\right]=6.6\%$
in the 2-BRE, while uncorrelated product is four orders of magnitude
smaller $P\left[J_{max}(5)_{0}\right]P\left[J_{max}(6)_{0}\right]\dots P\left[J_{max}(10)_{0}\right]=2.1\cdot10^{-4}\%.$
In Tab.\,\ref{tab:onejc} we show the total weighted correlation
which for a general set of events $J,J',J'',J'''\dots$ is defined
as \begin{equation}
\mathcal{C}\left[J,J',J'',\dots\right]=\frac{\log\left(P\left[J\right]P\left[J'\right]P\left[J''\right]\dots\right)}{\log\left(P\left[J,J',J'',\dots\right]\right)}-1.\label{eq:corrent}\end{equation}
 If ground states in $k$ systems with different masses are not correlated
then $\mathcal{C}=0$. While in the case of full correlation the ratio
of logarithms is approximately equal to $k$ and $\mathcal{C}=k-1$,
meaning that the g.s spin in one system is sufficient to predict g.s.
spins in all other $k-1$ different mass-systems. 

The situation with $n$-body forces is more complex as coherent higher
order terms appear. For the 3-BRE $\langle H^{(2)}\rangle_{J}=(\tilde{V}'N+\tilde{V})J^{2}$
and the moment inertia is particle-number dependent. However, the
3-BRE is expected to be similar to the 2-BRE where the g.s. spin statistics
is dictated primarily by the sign of the moment of inertia which equally
favors both $J_{min}$ and $J_{max}.$ Although the effect is reduced
by incoherent interaction components the preponderance of $J_{min}$
and $J_{max}$ is seen in Fig.\,\ref{fig:pairing} and Tab.\,\ref{tab:onejc}.
The $N$-dependence reduces the amount of correlation between systems
of different masses, see Tab.\,\ref{tab:onejc}. However, the variations
of $N$ which is relatively large and positive are usually insufficient
to change the sign in the moment of inertia and therefore correlations
between systems of different $N$ are strong.

The angular momentum dependent part in the 4-body forces is given
by a more complicated expression with four possible, normally distributed,
interaction terms $\langle H^{(2)}\rangle_{J}=(\tilde{V_{1}''N^{2}+}\tilde{V}_{1}'N+\tilde{V_{1}})J^{2}+\tilde{V_{2}}J^{4}.$
The coefficients $\tilde{V}_{2},\tilde{V}_{1},\tilde{V}_{1}',$ and
$\tilde{V_{1}}''$ are given by correlated normal distributions but
because of geometric complexity in the recoupling coefficients and
a large number of 4-body interaction parameters these correlations
are small. The $P\left[\left(J_{max}\right)_{0}\right]$ is expected
to be significantly reduced, with more chances going to the $P\left[\left(J_{min}\right)_{0}\right]$
which is always a local minimum for non-negative $J^{2}$, Tab.\,\ref{tab:onejc}
and Fig.\,\ref{fig:j19N87}. Only for a large particle number the
term $V_{1}''N^{2}J^{2}$ in the Hamiltonian seems to dominate enhancing
the $P\left[\left(J_{max}\right)_{0}\right],$ Tab.\,\ref{tab:onejc}.
Because of the strong $N$-dependence in the moment of inertia and
presence of the centrifugal distortion $\tilde{V}_{2}$ for $n=4$
there is much less correlation between g.s. spins for systems with
different masses, Tab.\,\ref{tab:onejc}.

\begin{table}
\begin{tabular}{|c||c|c||c|c||c|c|}
\hline 
 & \multicolumn{2}{c||}{2-BRE} & \multicolumn{2}{c||}{3-BRE} & \multicolumn{2}{c|}{4-BRE}\tabularnewline
\hline 
N  & $J_{min}$  & $J_{max}$  & $J_{min}$  & $J_{max}$  & $J_{min}$  & $J_{max}$\tabularnewline
\hline
\hline 
5  & 16.0  & 10.1  & 36.3  & 2.9  & 7.7  & 0.2\tabularnewline
\hline 
6  & 52.3  & 10.5  & 66.4  & 3.1  & 83.0  & 0.0\tabularnewline
\hline 
7  & 12.4  & 11.8  & 39.1  & 4.8  & 33.0  & 0.5\tabularnewline
\hline 
8  & 42.7  & 12.1  & 63.2  & 5.0  & 84.3  & 1.1\tabularnewline
\hline 
9  & 9.5  & 12.3  & 31.1  & 6.5  & 33.7  & 2.3\tabularnewline
\hline 
10  & 31.2  & 11.5  & 48.6  & 7.1  & 65.5  & 2.6\tabularnewline
\hline
\hline 
$\mathcal{C}$  & 1.883  & 3.806  & 1.560  & 3.266  & 0.435  & 0.000\tabularnewline
\hline
\end{tabular}

\caption{Summary of g.s. statistics for minimum and maximum spin, and correlations
across different mass numbers $N$ for a single $j=19/2$ valence
space with 2,3, and 4-body random interactions. The first six rows
correspond to $P[J(N)]$ expressed in percent. The lowest row shows
correlation $\mathcal{C}$, for all particle numbers $N=5,6,7,8,9,$
and 10. Columns reflect the type of ensemble and $J_{min}$ or $J_{max}.$\label{tab:onejc}}

\end{table}

The prevailing spin sequence of low-lying states provides yet another
information about the coherent symmetry structure of interactions.
The chances to find g.s. with spin $J_{0}=0$ followed by the first
excited state with $J_{1}=2$ and then by the second excited state
$J_{2}=4$ are high. In Tab.\,\ref{tab:Sequence} this probability
$P\left[0_{0},2_{1},4_{2}\right]$ is shown along with the average
ratio of the excitation energies for the $2_{1}$ and $4_{2}$ states.
The typical numbers between 2 and 10 \% reflect extremely high probability
compared to the chances of finding this sequence in a random list
of spins. However, the term {}``sequence'' must be used with reservations,
the weighted correlational entropy \eqref{eq:corrent} between the
joint $P\left[0_{0},2_{1},4_{2}\right]$ and the product of independent
\textbf{$P\left[0_{0}\right]$, $P\left[2_{1}\right]$} and \textbf{$P\left[4_{2}\right]$}
is only about $\mathcal{C}\left[0_{0},2_{1},4_{2}\right]\approx0.4$
in all cases. The number is comparable to unity showing definite correlations,
but they are not strong and the high probability to observe the sequence
comes partially from the independently high chances of finding low-lying
states with angular momenta 0, 2 and 4. These states do not always
form a rotational band which would lead to the ratio of excitation
energies $E_{1}/E_{2}=0.3$. It is likely that the members of the
ground state band are often higher in energy and further work is needed
to identify them.

\begin{table}
\begin{tabular}{|c|c|c|c|c|c|c|}
\hline 
 & \multicolumn{2}{c|}{2-BRE} & \multicolumn{2}{c|}{3-BRE} & \multicolumn{2}{c|}{4-BRE}\tabularnewline
\hline
 & $P$  & $E_{1}/E_{2}$  & $P$  & $E_{1}/E_{2}$  & $P$  & $E_{1}/E_{2}$\tabularnewline
\hline
\hline 
6  & 3.7(2)  & 0.55  & 4.2(2){*}  & 0.69  & 4.4(7){*}  & 0.69\tabularnewline
\hline
8  & 4.2(2){*}  & 0.59  & 5.5(2)  & 0.67  & 7.4(11)  & 0.75\tabularnewline
\hline
10  & 2.1(1){*}  & 0.72  & 5.2(2)  & 0.69  & 7.3(10)  & 0.62\tabularnewline
\hline
\end{tabular}

\caption{Probability of finding the three lowest states as a sequence 0,2,4,
$P\left[0_{0},2_{1},4_{2}\right],$ labeled as $P$, expressed in
percent, and the ratio of excitation energies between $2_{1}$ and
$4_{2}$ states. In all cases the sequence 0,2,4 in the most likely
g.s. sequence except for those marked with {*}. \label{tab:Sequence}}

\end{table}

\begin{figure}
\includegraphics[width=3.4in]{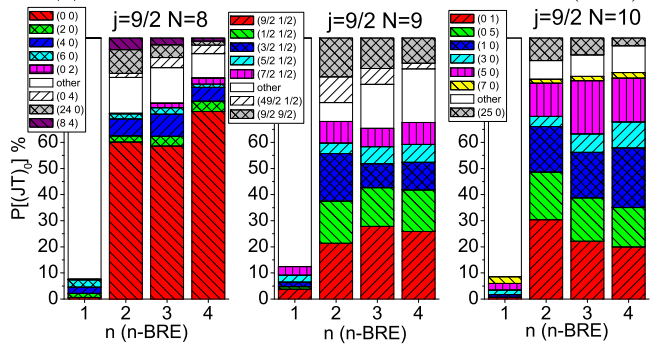}\caption{Statistics of g.s. quantum numbers labeled as $(J\, T)$ for isospin
$1/2$ fermions on a single $j=9/2$ level. The panels from left to
right correspond to $N=8,$ 9, and 10. The $n=1$ is equiprobable
distribution. Notations are similar to Fig.\,\ref{fig:j19N87}.\label{fig:j1T}}

\end{figure}

We conclude this work by showing the statistics of g.s. quantum numbers
for systems with the rotational and isospin symmetries in Fig.\,\ref{fig:j1T}.
The preponderance of the symmetric g.s. with $(J\, T)=(0\,0)$ is
robust for every particle number divisible by 4. Although geometrically
$(0\,0)$ state is possible in $j=19/2\,\, N=10$ it does not appear
as g.s. The possibility of $\alpha$-type correlations is to be investigated
in the future. The g.s. in the $N=9$ system can be explained as a
single particle coupled to the $N=8$ core, thus $T_{0}=1/2$ is favored
along with $(9/2\,\,9/2)$ and $(49/2\,\,1/2),$ the maximum isospin
or spin states. For $N=10$ the $P\left[(J\, T)_{0}\right]$ is only
substantial when either $J=0$ or $T=0$ with clear preference to
the quantum numbers of two coupled identical $j=9/2$ fermions ($J+T$
is odd). The preponderance of the minimum or maximum in either spin
or isospin appears in all cases. Probability to find g.s. with $J_{max}$
or $T_{max}$ diminish with higher $n.$ The overall picture is in
qualitative agreement with the hypothesis discussed above: that the
powers of $J^{2}$ and $T^{2}$ operators appear with random sign
as property of the coherent interaction components in each realization.
The mixed terms, such as $J^{2}T^{2}$ do not single out a particular
$(J\: T)$ pair. In relation to a more complex shell model studies
it is interesting to mention a $p$-shell ($j=3/2$ and $j=1/2)$
odd-odd $^{10}$B case, where results are similar to the above study
favoring g.s. of coupled two-particle quantum numbers. For a degenerate
$p$ levels $P\left[(1^{+}0)_{0}\right]=19.1(4)\%$, $P\left[(3^{+}0)_{0}\right]=36.6(6)\%$,
and $P\left[(0^{+}1)_{0}\right]=21.3(5)\%$ for 2-BRE; the same numbers
are 25.6(5), 30.5(6), 24.5(5) for 3-BRE; and 30.3(6), 29.9(6), 23.2(5)
for 4-BRE, respectively. The chances of $(J^{\pi}T)_{0}$ being $(1^{+}0)$
grow with $n$, although the statistics has little to do with the
realization chosen by nature \citep{Navratil:2002,Pieper:2001}. Further
exploration of more complex models will be reported elsewhere. 

To summarize, in this work the ensembles of fermions interacting randomly
with symmetry conserving many-body forces have been considered. The
preponderance of symmetry dominated ground state was observed. The
effect is generally stronger for the higher $n$-body forces. The
possibility of correlated paired ground state is discussed and a surprising
enhancement of pairing with three-body forces is observed. An explanation
based on the coherent components of interaction is suggested, which
qualitatively describes the results. The strong correlations between
systems of different particle-number, mass dependence of probabilities,
and correlated sequences of states support this theoretical hypothesis.
This work opens many future avenues for investigation, studies and
ideas \citep{Johnson:1998,Mulhall:2000,Papenbrock:2004,Zuker:2003}
from random two-body ensembles can be extended. The onset of coherence:
pairing, in isovector or isoscalar form, $\alpha$-particle four-body
clustering, shape properties and vibrations are all interesting and
important questions for future investigations. 

The author is grateful to V. Zelevinsky, D. Mulhall, and V. Abramkina
for motivating discussions. Support from the U. S. Department of Energy,
grant DE-FG02-92ER40750 is acknowledged. 

\bibliographystyle{apsrev} \bibliographystyle{apsrev}
\bibliography{nbrandom}

\end{document}